\documentclass[twocolumn,floatfix,superscriptaddress]{revtex4-1}
\usepackage{dcolumn}
\usepackage{bm}
\usepackage{graphicx}
\usepackage{amsmath,amssymb}
\usepackage{amsthm}
\usepackage{float}
\usepackage{physics}
\usepackage{color}
\usepackage[colorlinks=true,allcolors=blue]{hyperref}
\begin{document}

\title{Holographic energy loss near critical temperature in an anisotropic background}
\date{\today  \hspace{1ex}}
\author{Qi Zhou}
\email{qizhou@mails.ccnu.edu.cn}
\affiliation{Key Laboratory of Quark \& Lepton Physics (MOE) and Institute of Particle Physics,Central China Normal University, Wuhan 430079, China}

\author{Ben-Wei Zhang}
\email{bwzhang@mail.ccnu.edu.cn}
\affiliation{Key Laboratory of Quark \& Lepton Physics (MOE) and Institute of Particle Physics,Central China Normal University, Wuhan 430079, China}

\begin{abstract}
We study the energy loss of a quark moving in a strongly coupled QGP under the influence of anisotropy. The heavy quark drag force, diffusion coefficient, and jet quenching parameter are calculated using the Einstein-Maxwell-dilaton model, where the anisotropic background is characterized by an arbitrary dynamical parameter $A$.
Our findings indicate that as the anisotropic factor $A$ increases, the drag force and jet quenching parameter both increase, while the diffusion coefficient decreases. Additionally, we observe that the energy loss becomes more significant when the quark moves perpendicular to the anisotropy direction in the transverse plane.
The enhancement of the rescaled jet quenching parameters near critical temperature $T_c$, as well as drag forces for a fast-moving heavy quark is observed, which presents one of the typical features of QCD phase transition.
\end{abstract}

\maketitle

\section{Introduction}

The heavy-ion collisions (HICs) experiments at the Relativistic Heavy Ion Collider (RHIC) and the Large Hadron Collider (LHC) are believed to create almost the most perfect fluid Quark Gluon Plasma (QGP)~\cite{BRAHMS:2004adc,PHENIX:2004vcz,PHOBOS:2004zne,STAR:2005gfr}. This provides a novel window for studying the physics of Quantum Chromodynamics (QCD) at a strongly coupled regime.
Since the properties of a strongly coupled system cannot be reliably calculated directly by perturbative techniques, one has to resort to some nonperturbative approaches to overcome the challenges.

The AdS/CFT correspondence, initially proposed by Maldacena in 1997, makes a conjecture that the large $N_c$ limits of certain conformal field
theories in $d$-dimensions can be described in terms of string theory on the product of $(d+1)$-dimensional Anti-de Sitter space with a compact manifold~\cite{Maldacena:1997re,Witten:1998qj,Gubser:1998bc}.
Following the efforts of pioneers, the correspondence is introduced to handle problems in gauge theory at the strongly coupled scenario~\cite{Witten:1998zw,Aharony:1999ti,Casalderrey-Solana:2011dxg}. Especially, since the QCD is a multiple scales theory, finding a gravity dual to all scales of QCD is one of the essential aims of the AdS/CFT correspondence.
Although the precise gravity dual to QCD is still unknown, the $\mathcal{N} = 4$ supersymmetric Yang-Mills (SYM) and QCD may share the same qualitative features at finite temperature, which means one could capture the physics of strong coupled QCD by deformed $AdS_5$~\cite{Erlich:2009me,Peet:1998wn,Gursoy:2010fj}.
One of the significant achievements of AdS/CFT correspondence is the computation of the ratio of shear viscosity over to entropy density of the QGP, which is $1/4\pi$~\cite{Policastro:2001yc}, a simple universal value on the gravity side~\cite{Buchel:2003tz}. 
Besides, plenty of real-time dynamical quantities were computed on the weakly coupled gravity side with top-down~\cite{Polchinski:2000uf,
Karch:2002sh,Sakai:2004cn} and bottom-up~\cite{Gubser:2008ny,Gursoy:2007cb,Galow:2009kw} holographic QCD models, such as hydrodynamic transport coefficients~\cite{Herzog:2002fn,Benincasa:2006ei,Baier:2007ix,Mas:2007ng,Natsuume:2007ty}, energy loss of energetic parton travelling through the QGP~\cite{Herzog:2006gh,Gubser:2006bz,Ficnar:2013qxa,Liu:2006ug,BitaghsirFadafan:2017tci}, transverse momentum broadening~\cite{Casalderrey-Solana:2007ahi,Gubser:2006nz}, the thermal photon and di-lepton production rates~\cite{Caron-Huot:2006pee,PhysRevD.86.026003,Wu:2013qja,Patino:2012py} and so on~\cite{Zhu:2019igg,Rajagopal:2016uip,Zhu:2021vkj,Zhou:2021nbp}.

The QGP created during the experiments in HICs, is believed to be anisotropic both in momentum and coordinate space for a short time~\cite{Strickland:2013uga}. 
Roughly speaking, the pressures of the QGP along the transverse direction may be larger than the pressure along the beam direction at a very earlier time, due to the rapid expansion along the beam direction.
It is noticed that only the holographic QCD models with anisotropy succeeded in attempting to reproduce energy dependence of the total multiplicity of experiments in HICs~\cite{Kiritsis:2011yn,Arefeva:2013fia,Arefeva:2014vjl}.
With gauge gravity duality, the anisotropic geometries have been investigated to understand the properties of the QGP for a long time. The neutral spatial anisotropic black brane solution was found originally at zero temperature~\cite{Azeyanagi:2009pr} and soon at nonzero temperature~\cite{Mateos:2011tv,Mateos:2011ix}.
Furthermore, many other interesting constructions to this anisotropic system have been developed from different groups~\cite{Cheng:2014qia,Cheng:2014sxa,Banks:2015aca,Avila:2016mno,Giataganas:2017koz,Itsios:2018hff,Li:2022hka}.
Besides, the strong magnetic field also plays an important role in HICs and is also a source of anisotropy~\cite{DHoker:2009mmn,Dudal:2014jfa,Mamo:2015dea,Li:2016gfn}.
Most of the holographic work on anisotropic systems currently focuses on the systems with a magnetic field and the systems with spatial anisotropy, corresponding to the strong magnetic field created during HICs and the earlier time anisotropic phase of QGP produced in HICs.
Although the anisotropy in such models may be different from the real QGP, it is expected that this kind of effort could help to reveal some intrinsic features of this plasma~\cite{Rougemont:2020had,Zhang:2018mqt,Reiten:2019fta,Zhang:2018pyr,Rajagopal:2015roa,Zhang:2019jfq,Finazzo:2016mhm}.

Further, to locate the critical point on~$(\mu,T)$-plane and probe the properties around the critical point, lots of efforts have been devoted by different groups based on isotropic Einstein-Maxwell-dilaton (EMD) models~\cite{PhysRevD.83.086005,PhysRevD.84.126014,Gursoy:2007er,Ballon-Bayona:2017sxa,He:2013qq,Yang:2014bqa} and EMD models with magnetic fields where the anisotropy is introduced by an external field~\cite{Gursoy:2021vpu,Rougemont:2015oea,Bohra:2019ebj,He:2020fdi}.
Recently, the authors of \cite{Arefeva:2018hyo,Arefeva:2020byn} proposed a new version of the EMD model, where the anisotropy is introduced at one spatial direction in metric.
As we mentioned before, the metric ansatz in this model can accurately reproduce the energy dependence of the total particle multiplicity, which is one of our motivations for studying the energy loss near $T_c$ within this system.
It is illuminating to conduct an investigation on the energy loss of an energetic parton in the presence of anisotropy with this new anisotropic bottom-up QCD system.
Since the EMD model is designed to mimic the QCD deconfinement phase transition, it is also of great interest to utilize this anisotropic EMD model to study the propagation of a quark around critical temperature $T_c$.
With much attention having been attracted by the recent BES program in HICs, we hope our study can provide some insights into a better understanding of the real-time dynamical properties around QCD critical point.

This paper is organized as follows. In Sec.~\ref{chapter:background}, we briefly introduce the EMD model with the spatial anisotropic background~\cite{Arefeva:2020byn}.
In  Sec.~\ref{chapter:dragforce} we derive the drag force of heavy quark energy loss when passing through the QGP with the classic trailing string model.
In  Sec.~\ref{chapter:diffusion} we compute non-relativistic diffusion parameters by using Einstein relation together with the results of  Sec.~\ref{chapter:dragforce}.
And the numerical results of jet quenching parameters are discussed in  Sec.~\ref{chapter:qhat}. In the end, we present a short summary in Sec.~\ref{chapter:summary}.

\section{The EMD model}\label{chapter:background}
The EMD system with anisotropy has been studied by the authors of~\cite{Arefeva:2020byn}.
In this section, we briefly review this anisotropic holographic model starting from the Einstein-dilaton-two-Maxwell action,
\begin{multline}
    S=\int \frac{\mathrm{d}^{5} x}{16 \pi G_{5}} \sqrt{-g}\\
    \times\left[R-\frac{f_{1}(\phi)}{4} F_{(1)}^{2}-\frac{f_{2}(\phi)}{4} F_{(2)}^{2}-\frac{1}{2} \partial_{\mu} \phi \partial^{\mu} \phi-V(\phi)\right]
\end{multline}
where $F_{(1)}$ and $F_{(2)}$ are the field strength tensors of the two $U(1)$ gauge fields introduced to provide for the chemical potential and the anisotropy respectively, $\phi$ is the dilaton field and $V(\phi)$ denotes the dilaton potential. And $f_1(\phi)$ and $f_2(\phi)$ are the gauge kinetic functions representing the coupling with the two $U(1)$ gauge fields respectively.

For a holographic description of the hot and dense anisotropic QGP, one possible version of the anisotropic metric ansatz is employed as
\begin{multline}
\label{metric1}
d s^{2}=\frac{L^{2} b(z)}{z^{2}}\\
\times \left[-g(z) \mathrm{d} t^{2}+\mathrm{d} x^{2}+z^{2-\frac{2}{A}}\left(\mathrm{d} y_{1}^{2}+\mathrm{d} y_{2}^{2}\right)+\frac{\mathrm{d} z^{2}}{g(z)}\right]
\end{multline}
where $L$ gives the $AdS$-radius, $b(z)=\mathrm{e}^{2\mathcal{A}(z)}$ denotes the warp factor, $g(z)$ stands for the blackening function. 
Following~\cite{Li:2017tdz,Arefeva:2018hyo}, the function $\mathcal{A}=-a\mathrm{ln}(bz^2 + 1)$ is chosen for a light quark system, and the function $f_{1}=\mathrm{e}^{-cz^2-\mathcal{A}(z)}z^{-2+\frac{2}{A}}$ is determined to reproduce the Regge spectrum.
Since there is rotational invariance in the $y_1y_2$-direction, we denote the $x$-axis as the direction of anisotropy. 

The arbitrary dynamical parameter $A$ measures the degree of anisotropy and Lorentz symmetry violation in $y_1y_2$-plane. 
A relativistic jet parton is focused on in our study, and we intend to introduce a slight break in symmetry by setting the value of $A$ very close to unit.
As in~\cite{Arefeva:2020byn}, the authors found that continually increasing the value of $A$ had a dramatic impact on the thermal properties of the system.
Therefore we chose to constrain our calculations at zero chemical potential using slight anisotropy cases with $A$ values of 1.01, 1.02, and 1.03 for convenience.

In the following calculations, we set the $AdS$ radius $L$ to be one for convenience. The solution for the blackening function may be obtained in
\begin{multline}
g(z)=1-\frac{\int_0^z(1+b\xi^2)^{3a}\xi^{1+\frac{2}{A}}\mathrm{d}\xi}{\int_0^{z_h}(1+b\xi^2)^{3a}\xi^{1+\frac{2}{A}}\mathrm{d}\xi}\\+\frac{2\mu^2c}{L^2(1-\mathrm{e}^{c z_h^2})^2}\int^{z}_{0}\mathrm{e}^{c\xi^2}(1+b\xi^2)^{3a}\xi^{1+\frac{2}{v}}\mathrm{d}\xi\\
\times\left[1-\frac{\int_0^z(1+b\xi^2)^{3a}\xi^{1+\frac{2}{A}}\mathrm{d}\xi}{\int_0^{z_h}(1+b\xi^2)^{3a}\xi^{1+\frac{2}{A}}\mathrm{d}\xi}\frac{\int_0^{z_h}\mathrm{e}^{c\xi^2}(1+b\xi^2)^{3a}\xi^{1+\frac{2}{A}}\mathrm{d}\xi}{\int_0^{z}\mathrm{e}^{c\xi^2}(1+b\xi^2)^{3a}\xi^{1+\frac{2}{A}}\mathrm{d}\xi}\right].
\end{multline}

Calculating the derivative of the blackening function, the temperature is parameterized by $z_h$,
\begin{multline}
T =\frac{\left|g^{\prime}(z)\right|}{4 \pi}\bigg|_{z=z_h}\\
=\frac{1}{4\pi}\left| -\frac{(1+b z_h^2)^{3a}z_h^{1+\frac{2}{A}}}{\int_{0}^{z_h}(1+b\xi^2)^{3a}\xi^{1+\frac{2}{A}}\mathrm{d}\xi}\left[1-\frac{2\mu^2c \mathrm{e}^{2c z_h^2}}{L^2(1-\mathrm{e}^{c z_h^2})^2}\right.\right.\\
\times\left(1-\mathrm{e}^{-c z_h^2}\frac{\int_0^{z_h}\mathrm{e}^{c\xi^2}(1+b\xi^2)^{3a}\xi^{1+\frac{2}{A}}\mathrm{d}\xi}{\int_0^{z_h}(1+b\xi^2)^{3a}\xi^{1+\frac{2}{A}}\mathrm{d}\xi}\right)\\
\times\left.\left.\int^{z_h}_{0}\left(1+b\xi^2\right)^{3a}\xi^{1+\frac{2}{A}}\mathrm{d}\xi\right]\right|,
\end{multline}
where $z_h$ denotes the location of the horizon.

And the dilaton field $\phi(z)$ reads
\begin{equation}
\phi(z)=\int^{z}_{z_0}\mathrm{d}\xi\times \frac{2\sqrt{A-1+\left[2(A-1)+9aA^2\right]b\xi^2+K}}{(1+b\xi^2)A\xi}\\
\end{equation}
and
\begin{equation}
    K=\left[A-1+3a(1+2a)A^2\right]b^2\xi^4.
\end{equation}

There are no divergences in the isotropic case $A=1$ for the dilaton field, but in the anisotropic cases, the dilaton field has a logarithmic divergence with $\phi(z)\approx\frac{2\sqrt{A-1}}{A}\mathrm{Log}(\frac{z}{z_h})$. It is proposed in~\cite{Arefeva:2020byn,slepov2021way} that a sufficiently small boundary condition point $z_0$ should reproduce the proper behavior of the scalar field.
In this paper, we take $a = 4.046, \  b = 0.01613, \  c = 0.227$ to be compatible with results in the isotropic case~\cite{Li:2017tdz}, where the critical temperature is around $T_c = 157.8$ MeV.

\section{Drag force}
\label{chapter:dragforce}
In small momentum transfer limit, the multiple scattering of heavy quarks with thermal partons in the QGP can be treated as Brownian motion \cite{Cao:2011et,Moore:2004tg}, which can be described by the Langevin equation as,
\begin{equation}
\frac{\mathrm{d}p}{\mathrm{d}t}=-\eta_Dp + f_{drive}.
\end{equation}
When the heavy quark moves with a constant velocity $v$, the driving force $f_{drive}$ is equal to the drag force $f_{drag} = \eta_Dp$.

In gauge theory side, the heavy quark suffers a drag force and consequently loses its energy while traveling through the strongly coupled plasma. On gravity side, this process could be modeled by a trailing string~\cite{Herzog:2006gh,Gubser:2006bz}, and the drag force $f_{SYM}$ in isotropic SYM plasma with zero chemical potential is then given by
\begin{equation}
    \label{drag0}
    f_{SYM}=-\frac{\pi T^2 \sqrt{\lambda} }{2}\frac{v}{\sqrt{1-v^2}},
\end{equation}
where $\sqrt{\lambda}=\frac{L^2}{\alpha'}=\sqrt{g_{YM}N_{c}}$.
The energy loss of the heavy quark can be understood as the energy flow from the endpoint along the string towards the horizon of the world sheet.

We follow the argument in~\cite{Herzog:2006gh,Gubser:2006bz} to analyze the energy loss of a heavy quark in the anisotropic background. The drag forces are calculated near the critical temperature $T_c$, and the string dynamics are captured by the Nambu-Goto string world-sheet action
 \begin{equation}
    \label{nbaction}
     S=-\frac{1}{2\pi\alpha'}\int \mathrm{d}\tau \mathrm{d}\sigma \sqrt{-\mathrm{det} g_{\alpha\beta}},
 \end{equation}
 \begin{equation}
     g_{\alpha\beta}=\frac{\partial X^{\mu}}{\partial \sigma^{\alpha}}\frac{\partial X^{\nu}}{\partial \sigma^{\beta}}.
 \end{equation}
where $g_{\alpha\beta}$ is the induced metric, and $g_{\mu\nu}$ and $X_{\mu}$ are the brane metric and target space coordinates.

The trailing string corresponding to a quark moving on the boundary along the chosen direction $x_{p}(x_p=x,y_1,y_2)$ with a constant velocity $v$ has the usual parametrization
\begin{equation}
    \label{gauge}
    t=\tau, x_{p}=vt+\xi(z),z=\tau.
\end{equation}

Plugging static gauge Eq.~(\ref{gauge}) into the metric Eq.~(\ref{metric1}), we have
\begin{align}
    \label{inducemetric}
    \mathrm{d}s^2=g_{tt}\mathrm{d}t^2+g_{xx}\mathrm{d}x_p^2+g_{zz}\mathrm{d}z^2,&\\
    g_{tt}=\frac{-L^2 b(z) g(z)}{z^2},&\\
    g_{xx}=\frac{L^2 b(z)}{z^2 g(z)}|_{(x_p=x)},&\\
    g_{xx}=\frac{L^2 b(z)(z^{2-\frac{2}{A}})}{z^2 g(z)}|_{(x_p=y_1,y_2)},&\\
    g_{zz}=\frac{L^2 b(z)}{z^2g(z)}.
\end{align}

The Lagrangian density can be obtained from the Nambu-Goto action as
\begin{equation}
    \label{lag}
    \mathcal{L}=\sqrt{-g_{tt}g_{zz}-g_{zz}g_{xx}v^2-g_{tt}g_{xx}\xi'^2}
\end{equation}
The Lagrangian density does not depend on $\xi$ from Eq.~(\ref{lag}), which implies that the canonical
momentum is conserved,
\begin{equation}
    \Pi_{\xi}=\frac{\partial \mathcal{L}}{\partial\xi'}
    =\frac{-g_{tt}g_{xx}\xi'}{\sqrt{-g_{tt}g_{zz}-g_{zz}g_{xx}v^2-g_{tt}g_{xx}\xi'^2}}
\end{equation}

Then one can get
\begin{equation}
    \label{xx}
    \xi^{2}=\frac{-g_{zz}(g_{tt}+g_{xx}v^2)\Pi_{\xi}^{2}}{g_{tt}g_{xx}(g_{tt}g_{xx}+\Pi_{\xi}^{2})}
\end{equation}
Both the numerator and the denominator must change sign at the same location $z$ from
Eq.~(\ref{xx}). The critical point $z_c$ can be written as
\begin{equation}
    g_{tt}(z_{c})=-g_{xx}(z_{c})v^2,
\end{equation}
and
\begin{equation}
    \Pi^{2}_{\xi}=-g_{tt}(z_{c})g_{xx}(z_{c}).
\end{equation}

Finally, we obtain the drag force
\begin{equation}
\label{drag}
    f=-\frac{1}{2\pi\alpha'}\Pi_{\xi}=-\frac{1}{2\pi\alpha'}g_{xx}(z_{c})v.
\end{equation}

There are two different drag forces, $f^{v\parallel A}$ and $f^{v\perp A}$, for the anisotropy in background metric in Eq.~(\ref{metric1}).  To be specific, $f^{v\parallel A}$ stands for the drag force in parallel with the anisotropy direction, when the jet parton moves along the anisotropy direction. And $f^{v\perp A}$ denotes the drag force in parallel with its motion direction when the jet parton moving perpendicular to the anisotropy direction.
Plugging Eq.~(\ref{inducemetric}) into Eq.~(\ref{drag}), we have
\begin{align}
\label{dra1}
    f^{v\parallel A}&=-\frac{1}{2\pi\alpha'}g_{xx}(z_{c})v|_{x_p=x}\\
    &=-\frac{v}{2\pi\alpha'}\frac{b(z_{c})}{z_{c}^2},
\end{align}
and
\begin{align}
\label{dra2}
    f^{v\perp A}&=-\frac{1}{2\pi\alpha'}g_{xx}(z_{c})v|_{x_p=y_1}\\
    &=-\frac{v}{2\pi\alpha'}b(z_{c})z_{c}^{-\frac{2}{A}}.
\end{align}

\begin{figure}[hbtp]
    \includegraphics[width=8.1cm]{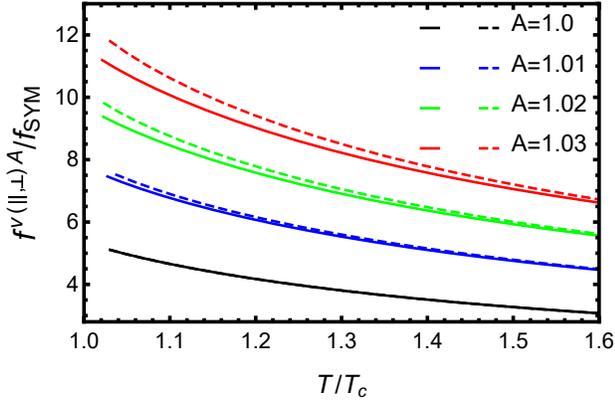}
    \caption{Perpendicular (dashed line) and parallel (solid line) drag force at lower speed~($v=0.6$) normalized by conformal limit as a function of the temperature for different values of the anisotropy factor $A$.}
    \label{figure:dragpara}
\end{figure}

\begin{figure}[hbtp]
    \includegraphics[width=8.1cm]{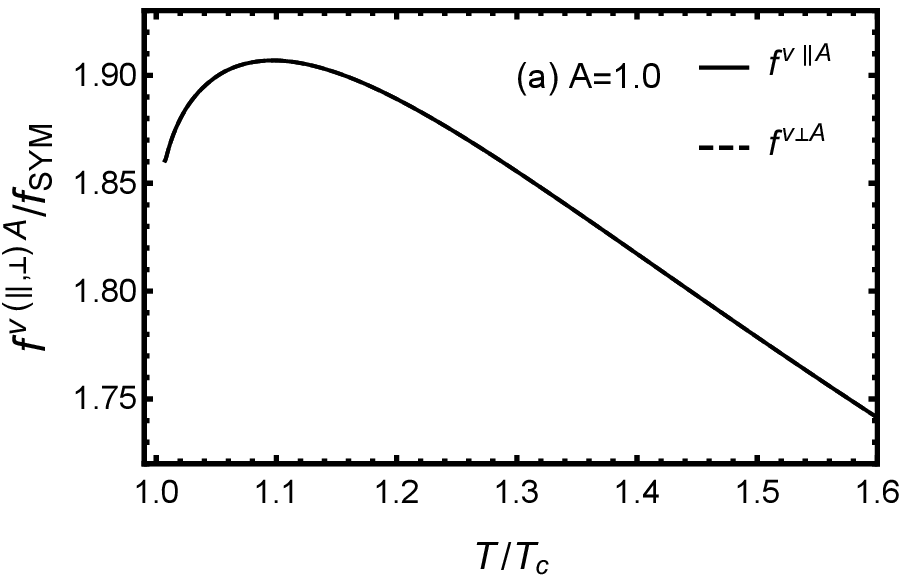}
    \includegraphics[width=8.1cm]{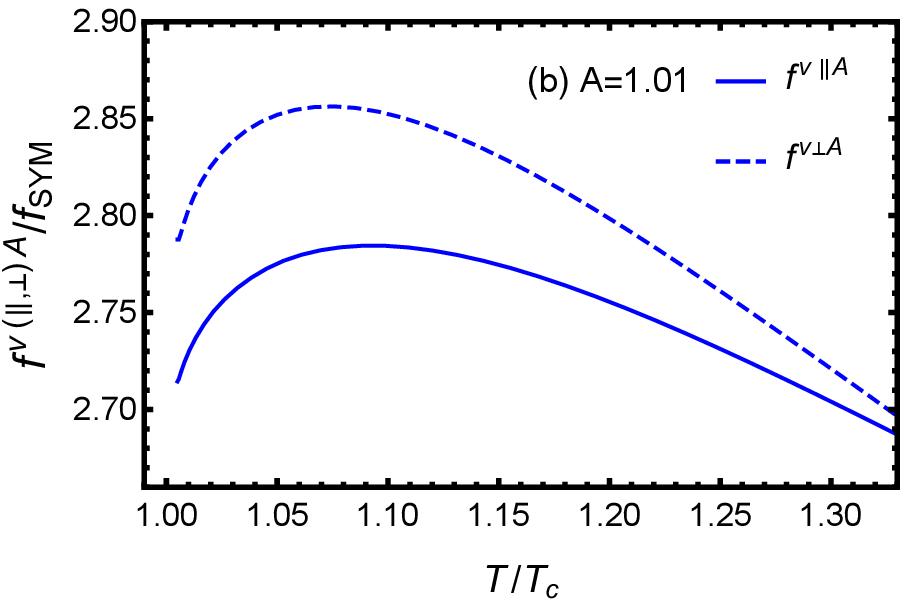}
    \includegraphics[width=8.1cm]{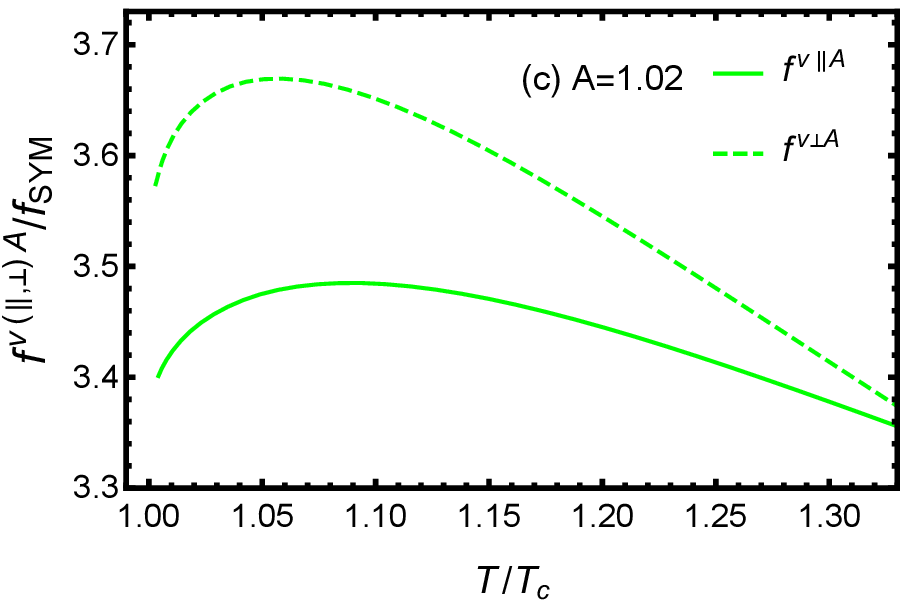}
    \includegraphics[width=8.1cm]{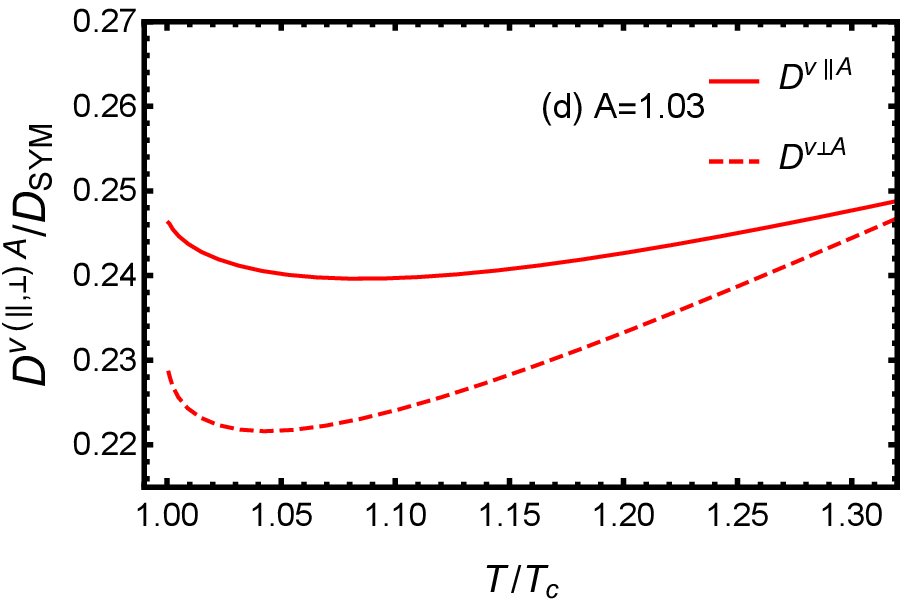}
    \caption{Perpendicular (dashed line) and parallel (solid line) drag force at higher speed~($v=0.96$) normalized by conformal limit as a function of the temperature for different values of the anisotropy factor $A$.}
    \label{figure:dragperp}
\end{figure}
The influence of spatial anisotropy on drag forces is illustrated in Fig.~\ref{figure:dragpara} and Fig.~\ref{figure:dragperp}, where the drag forces in anisotropic plasma are rescaled by the isotropic SYM result at zero chemical potential given in Eq.~(\ref{drag0}).  Fig.~\ref{figure:dragpara} shows, at lower speed~($v=0.6$) the drag force $f^{v\parallel A}$ always becomes larger with increasing anisotropic factor $A$.
A similar trend is also observed for drag forces perpendicular to anisotropy direction $f^{v\perp A}$. It is seen that the perpendicular direction drag force $f^{v\perp A}$ is larger than the parallel direction drag force $f^{v\parallel A}$ with the same anisotropic factor around critical temperature $T_c$.

At higher speed $v=0.96$, corresponding to a faster charm quark, the situation becomes more complicated as presented in Fig.~\ref{figure:dragperp}. Plots (a), (b), (c), and (d) in Fig.~\ref{figure:dragperp} present drag forces at different anisotropy $A=1$, $A=1.01$, $A=1.02$ and $A=1.03$ respectively. We find the drag force $f^{v\parallel A}$ always goes up with increasing anisotropic factor $A$. The perpendicular direction drag force $f^{v\perp A}$ is larger than the parallel direction drag force $f^{v\parallel A}$ around $T_c$. Furthermore, there is a peak near critical temperature $T_c$ when the velocity of quark ~($v=0.96$) is approaching the speed of light. The enhancement of energy loss around critical temperature $T_c$ is one of the typical features of QCD phase transition.
From Fig.~\ref{figure:dragpara} and Fig.~\ref{figure:dragperp},
we also find that the charm quark (with faster velocity and lighter mass) is more sensitive to properties of the anisotropy QGP than the bottom quark (with slower velocity and heavier mass) when they pass through the anisotropic plasma with a fixed initial energy $E_i$.

\section{Diffusion coefficient}\label{chapter:diffusion}
The diffusion coefficient, another important transport parameter of plasma, has been studied extensively at the RHIC and the LHC. It is of a general practice to utilize the Einstein-Maxwell system to study this transverse momentum broadening when heavy quark propagation in plasma~\cite{Zhu:2019ujc,Zhu:2021nbl}. The heavy quark transverse momentum diffusion constant $D$ in the strongly coupled $\mathcal{N} = 4$ supersymmetric Yang-Mills theory was first computed in~\cite{Casalderrey-Solana:2007ahi,Gubser:2006nz}, and then it was generalized to non-conformal theories in~\cite{Gursoy:2010aa}.
The Langevin dynamics of non-relativistic heavy quarks are completely determined by the momentum broadening $D$. The Einstein relation together with the expression of $\eta_D$ allows us to infer the value of $D$ for this strongly coupled anisotropic plasma.
The diffusion coefficient in the isotropic SYM theory \cite{Gubser:2006nz} is
\begin{equation}
    \label{dsym}
    D_{SYM}=\frac{T}{m}t_{SYM}=\frac{2}{\pi T \sqrt{\lambda}},
\end{equation}
where $t_{SYM} = \frac{1}{\eta_D}$ is the diffusion time.

From Eq.~(\ref{drag}) and Eq.~(\ref{dsym}), we obtain diffusion coefficient in anisotropic plasma normalized by isotropic SYM results as,
\begin{equation}
    \label{diff}
    \frac{D}{D_{SYM}}=\frac{\pi^2T^2}{g_{xx}(z_c)\sqrt{1-v^2}}|_{(x_p=x,y_1,y_2)}.
\end{equation}

Now there are also two different diffusion coefficients,  $D^{v\parallel A}$ and $D^{v\perp A}$, for the anisotropy in background metric Eq.~(\ref{metric1}).  $D^{v\parallel A}$ gives the diffusion coefficient when jet partons move along anisotropy direction, while $D^{v\perp A}$ gives the one when the jet parton moves perpendicular to anisotropy direction. Plugging Eq.~(\ref{inducemetric}) into Eq.~(\ref{diff}), we have
\begin{align}
    \frac{D^{v \parallel A}}{D_{SYM}}&=\frac{\pi^2T^2}{g_{xx}(z_c)\sqrt{1-v^2}}|_{(x_p=x)}\\
    &=\frac{\pi^2T^2}{b(z_c)z_{c}^{-2}\sqrt{1-v^2}}
\end{align}
and
\begin{align}
    \frac{D^{v \perp A}}{D_{SYM}}&=\frac{\pi^2T^2}{g_{xx}(z_c)\sqrt{1-v^2}}|_{(x_p=y_1)}\\
    &=\frac{\pi^2T^2}{b(z_c)z_{c}^{-\frac{2}{A}}\sqrt{1-v^2}}.
\end{align}

\begin{figure}[hbtp]
    \includegraphics[width=8.1cm]{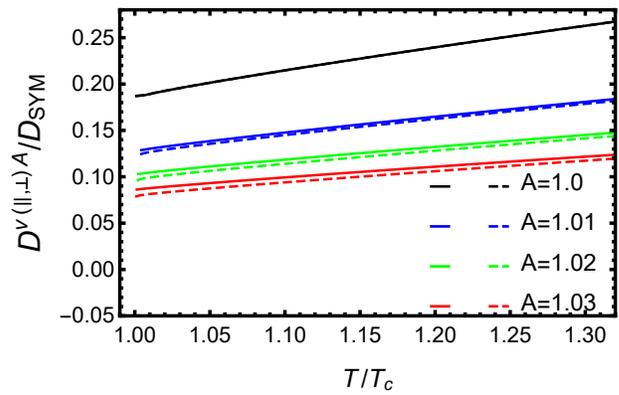}
    \caption{Perpendicular~(dashed line) and parallel~(solid line) diffusion constants at lower speed~($v=0.6$) normalized by conformal limit as a function of the temperature for different values of the anisotropy factor~$A$.}
    \label{figure:diffusionpara}
\end{figure}

\begin{figure}[hbtp]
    \includegraphics[width=8.1cm]{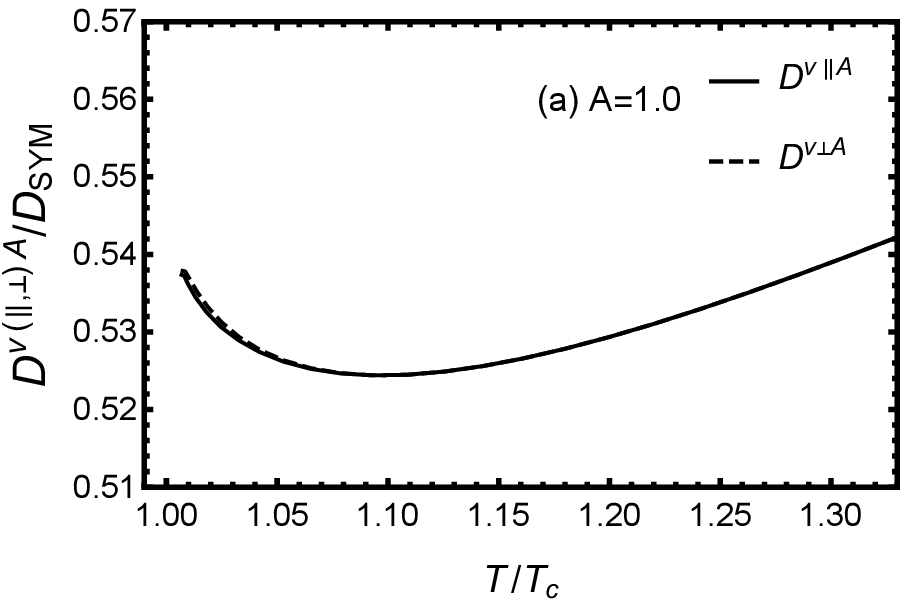}
    \includegraphics[width=8.1cm]{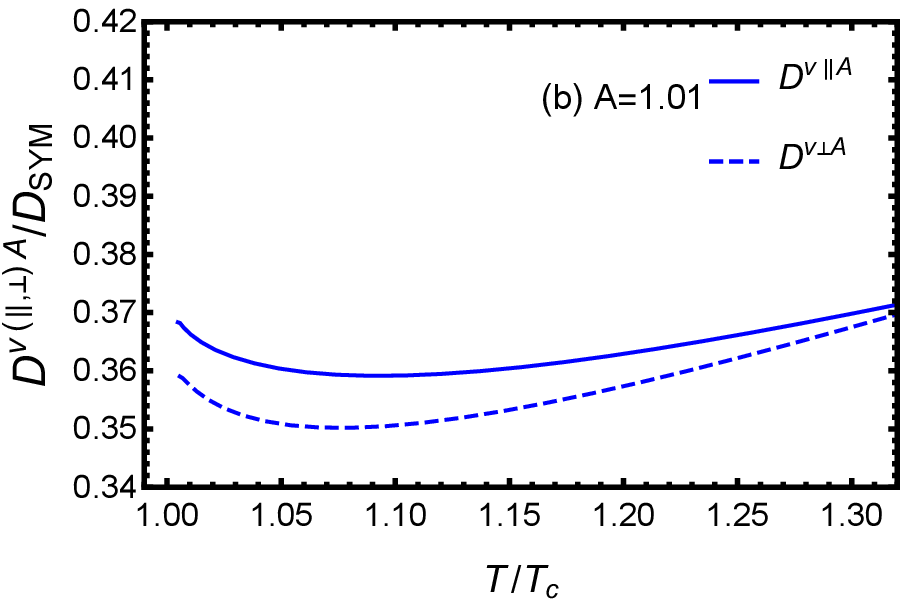}
    \includegraphics[width=8.1cm]{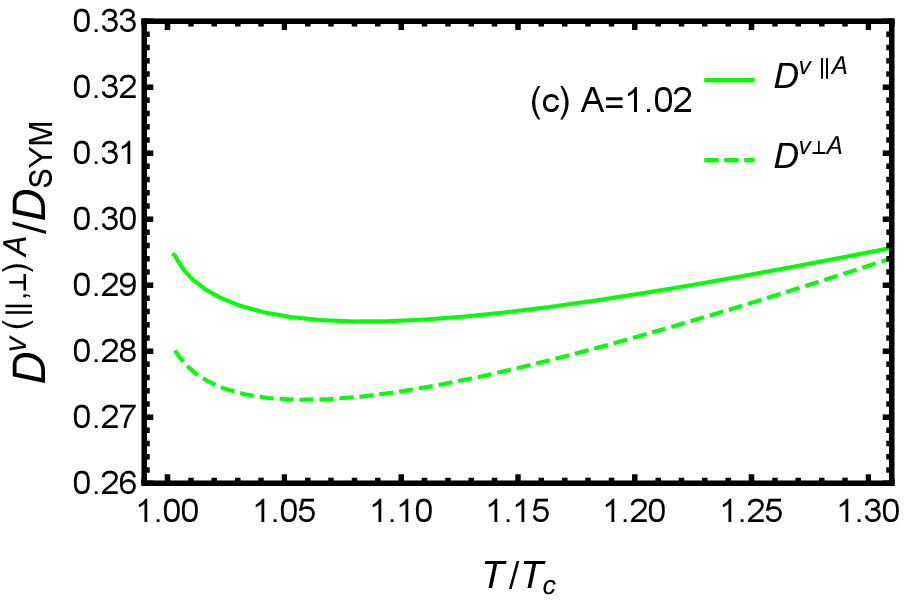}
    \includegraphics[width=8.1cm]{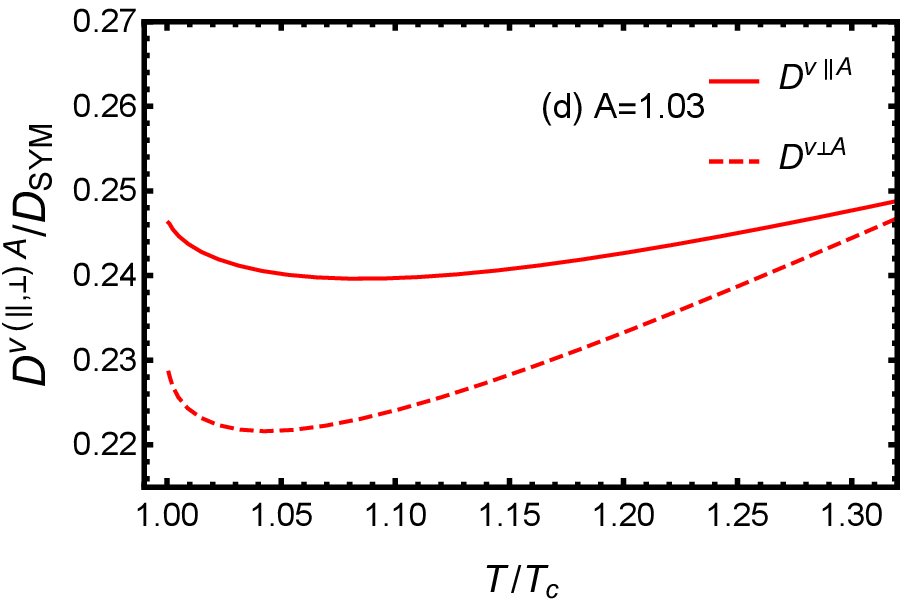}
    \caption{Perpendicular~(dashed line) and parallel~(solid line) diffusion constants at higher speed~($v=0.96$) normalized by conformal limit as a function of the temperature for different values of the anisotropy factor~$A$.}
    \label{figure:diffusionperp}
\end{figure}

The numerical results of the influences on diffusion constants $D$ from anisotropy factor are displayed in Fig.~\ref{figure:diffusionpara} and Fig.~\ref{figure:diffusionperp}, normalized by the isotropic SYM result at zero baryon density given in Eq.~(\ref{dsym}).
It is seen from Fig.~\ref{figure:diffusionpara} that, at lower speed~($v=0.6$) both $D^{v\parallel A}$  and $D^{v\perp A}$ suffer stronger
suppression with increasing anisotropic factor $A$. In addition, the perpendicular direction diffusion constant $D^{v\perp A}$ has stronger suppression than parallel direction diffusion constant $D^{v\parallel A}$ around critical temperature $T_c$.

Fig.~\ref{figure:diffusionperp} gives the results at a higher speed~($v=0.96$), and plots (a), (b), (c), and (d) in Fig.~\ref{figure:diffusionperp} present diffusion constants at different anisotropy $A=1$, $A=1.01$, $A=1.02$ and $A=1.03$ respectively. We find diffusion constant $D^{v\parallel A}$ goes down with increasing anisotropic factor $A$. It is also seen that the perpendicular direction diffusion constant $D^{v\perp A}$ has stronger suppression than parallel direction diffusion constant $D^{v\parallel A}$ around critical temperature $T_c$. One may observe the strongest suppression near critical temperature $T_c$ when the quark moves almost with the speed of light~($v=0.96$). All numerical results show the same trend that the energy loss in the perpendicular direction is larger than the one in the parallel direction.

\section{Jet transport parameter}\label{chapter:qhat}

In dual gravity theory,  a non-peturbative definition of jet transport coefficient $\hat{q}$ has been provided~\cite{Liu:2006ug}, based on the computation of light-like adjoined Wilson loops for $\mathcal{N}$ = 4 SYM plasma. It has been shown that the jet quenching parameter at an isotropic SYM plasma with zero chemical potential is
\begin{equation}
    \label{qhatsym}
    \hat{q}_{SYM}=\frac{\pi^{\frac{3}{2}}\sqrt{
    \lambda}T^{3}\Gamma(\frac{4}{3})}{\Gamma(\frac{5}{4})}.
\end{equation}
In this section, we discuss the jet quenching parameter $\hat{q}$ in the anisotropic background.
We follow the argument in~\cite{Liu:2006ug,Giataganas:2012zy} to study the jet quenching parameter of a light quark system in an anisotropic medium, in which the jet quenching parameter $\hat{q}$ is directly related to light-like adjoined Wilson loop~\cite{Liu:2006ug} as
\begin{equation}
  \langle W^{A}[\mathcal{C}] \rangle \approx \exp[-\frac{1}{4\sqrt{2}}\hat{q}L^{2}L^{-}],
\end{equation}
where $\mathcal{C}$ is a null-like rectangular Wilson loop formed by a quark-antiquark pair, $L$ gives the separated distance, and $L^{-}$ is the traveling distance along light-cone time duration.

Using the equations
\begin{equation}
    \langle W^A[\mathcal{C}] \rangle \approx \langle W^F[\mathcal{C}] \rangle^2
\end{equation}
and
\begin{equation}
   \langle  W^F[\mathcal{C}]  \rangle \approx \exp[-S_{I}],
\end{equation}
we obtain a general relation of jet quenching parameter
\begin{equation}
    \label{qhatraw}
    \hat{q}=8\sqrt{2}\frac{S_{I}}{L^{-}L^2}.
\end{equation}

To calculate the Wilson loop, we take advantage of the light-cone coordinates
\begin{equation}
    x_{+}=\frac{t+x_p}{\sqrt{2}}, \quad \quad x_{-}=\frac{t-x_p}{\sqrt{2}},
\end{equation}
where $x_p$ is chosen to be the direction of motion.

The metric Eq.~(\ref{metric1}) is then given by
\begin{align}
    \mathrm{d}s^2=
    G_{--}(\mathrm{d}x^{2}_{+}+\mathrm{d}x^{2}_{-})+G_{+-}\mathrm{d}x_{+}\mathrm{d}x_{-}&\\+G_{ii}\mathrm{d}x_{i}^{2}+G_{zz}\mathrm{d}z^2,&\\
    G_{--}=\frac{g_{tt}+g_{pp}}{2},&\\
    G_{+-}=\frac{g_{tt}-g_{pp}}{2},&\\
    G_{ii}=g_{ii}|_{(i=x,y_1,y_2)},&\\
    G_{zz}=g_{zz}.
\end{align}

Given the Wilson loop extending along the $x_k$ direction, we choose static gauge coordinates
\begin{equation}
   x_{-}=\tau, x_{k}=\sigma,u=u(\sigma).
\end{equation}
\begin{figure*}[hbtp]
    \centering
    \includegraphics[width=8.1cm]{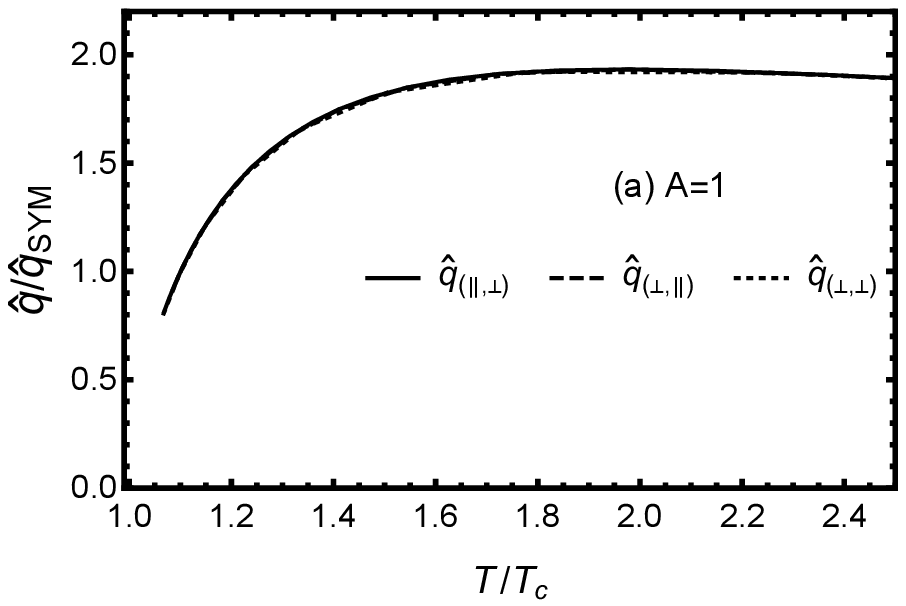}
    \includegraphics[width=8.1cm]{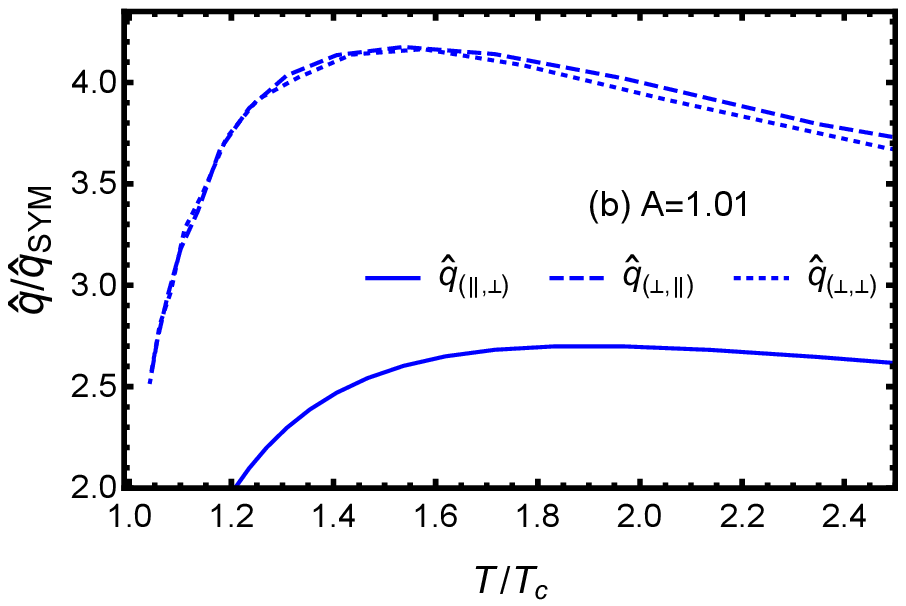}
    \includegraphics[width=8.1cm]{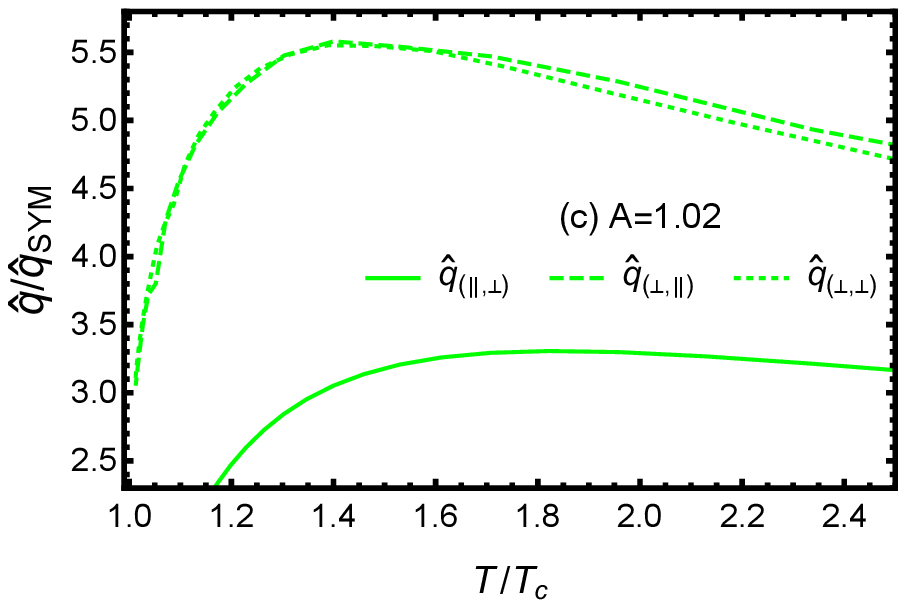}
    \includegraphics[width=8.1cm]{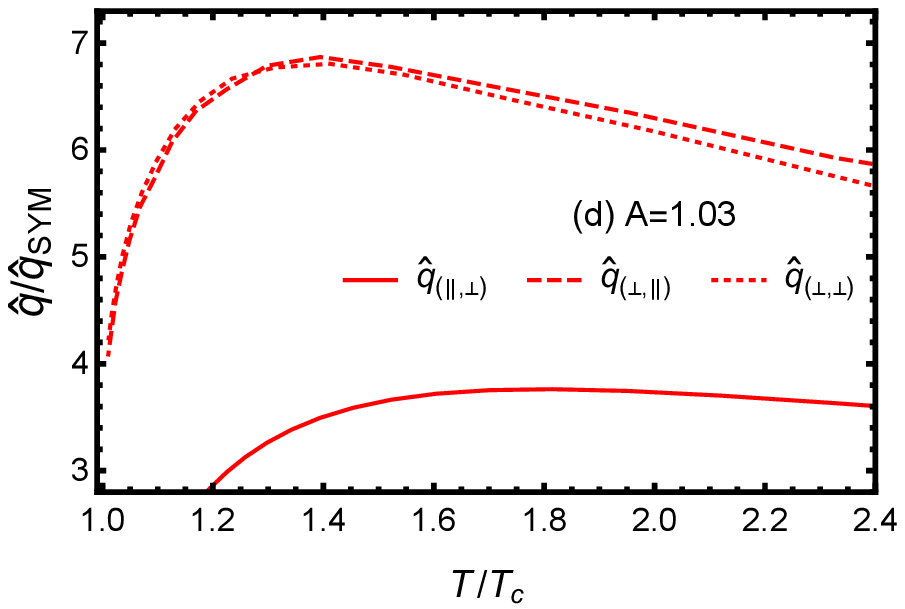}
    \caption{Motion parallel to anisotropy~(solid line) and perpendicular to anisotropy~(dashed and dotted line) jet quenching parameter normalized by conformal limit as a function of the temperature for different values of the anisotropy factor~$A$.}
    \label{figure:qhat}
\end{figure*}
The Nambu-Goto action Eq.~(\ref{nbaction}) can be given as
\begin{equation}
    \label{action2}
    S=\frac{L_{-}}{\pi\alpha'}\int^{\frac{L_{k}}{2}}_{0}\mathrm{d}\sigma\sqrt{G_{--}(G_{zz}z^{'2}+G_{kk})}.
\end{equation}

As action Eq.~(\ref{action2}) does not depend explicitly on $\sigma$ explicitly, we could have a conserved quantity $E$
\begin{equation}
    \frac{\partial \mathcal{L}}{\partial \Dot{z}}\Dot{z}-\mathcal{L}=E,
\end{equation}
resulting in
\begin{equation}
    \label{con1}
    \Dot{z}^{2}=\frac{(G_{kk}G_{--}-c^2)G_{kk}}{c^2G_{zz}}.
\end{equation}

Combining Eq.~(\ref{con1}) and Eq.~(\ref{action2}), we get
\begin{equation}
 S_{0}=\frac{L_{-}}{\pi\alpha'}  \int^{z_h}_{0}\mathrm{d}z\sqrt{G_{uu}G_{--}}.
\end{equation}

The total action is divergent and should be subtracted by the self-energy of the two free quarks part
\begin{equation}
\label{daction}
S-S_{0}=\frac{L_{-}}{\pi\alpha'}  \int^{z_h}_{0}\mathrm{d}z\sqrt{G_{zz}G_{--}}(\sqrt{\frac{G_{--}G_{kk}}{G_{--}G_{kk}-E^2}}-1)
\end{equation}

In our calculation, indices $p$ and $k$ here denote a chosen direction. Substituting Eq.~(\ref{daction}) into Eq.~(\ref{qhatraw}), we show
\begin{equation}
\label{jqp}
\begin{split}
\hat{q}_{(p,k)}&=\frac{\sqrt{2}}{\pi \alpha '}(\int^{z_{h}}_{0}\mathrm{d}z \frac{1}{G_{kk}}\sqrt{\frac{G_{zz}}{G_{--}}})^{-1}\\
&=\frac{\sqrt{2}}{\pi \alpha'}(\int^{z_{h}}_{0}\mathrm{d}z\frac{1}{g_{kk}}\sqrt{\frac{2g_{zz}}{g_{tt}+g_{pp}}})^{-1}
\end{split}
\end{equation}

Now we see there are three types of jet quenching parameters, $\hat{q}_{(\parallel,\perp)}$, $\hat{q}_{(\perp,\parallel)}$ and $\hat{q}_{(\perp,\perp)}$ for anisotropic background given in Eq.~(\ref{metric1}).  Here $\hat{q}_{(\parallel,\perp)}$ denotes the jet transport coefficient when energetic partons move along the anisotropy direction, and the momentum broadens perpendicular to anisotropy direction;
$\hat{q}_{(\perp,\parallel)}$ stands for the jet quenching parameter when energetic partons move perpendicular to the anisotropy direction, and the momentum broadening along anisotropy;  $\hat{q}_{(\perp,\perp)}$ gives the coefficient with fast parton moving perpendicular to anisotropy direction, and the momentum broadening perpendicular to anisotropy direction.
\begin{equation}
\label{jqp1}
\begin{split}
\hat{q}_{(\parallel,\perp)}&=\frac{\sqrt{2}}{\pi \alpha'}\left(\int^{z_{h}}_{0}\mathrm{d}z\frac{1}{g_{y_1y_1}}\sqrt{\frac{2g_{zz}}{g_{tt}+g_{xx}}}\right)^{-1}\\
&=\frac{\sqrt{2}}{\pi \alpha'}\left[\int^{z_{h}}_{0}\mathrm{d}z\left(b(z)z^{\frac{2}{A}}\sqrt{\frac{2}{g(z)(-f(z)+1)}}\right)\right]^{-1}
\end{split}
\end{equation}

\begin{equation}
\label{jqp2}
\begin{split}
\hat{q}_{(\perp,\parallel)}&=\frac{\sqrt{2}}{\pi \alpha'}\left(\int^{z_{h}}_{0}\mathrm{d}z\frac{1}{g_{xx}}\sqrt{\frac{2g_{zz}}{g_{tt}+g_{y_1y_1}}}\right)^{-1}\\
&=\frac{\sqrt{2}}{\pi \alpha'}\left[\int^{z_{h}}_{0}\mathrm{d}z\left(b(z)z^{-2}\sqrt{\frac{2}{g(z)(-f(z)+z^{2-\frac{2}{A}})}}\right)\right]^{-1}
\end{split}
\end{equation}

\begin{equation}
\label{jqp3}
\begin{split}
\hat{q}_{(\perp,\perp)}&=\frac{\sqrt{2}}{\pi \alpha'}\left(\int^{z_{h}}_{0}\mathrm{d}z\frac{1}{g_{y_1y_1}}\sqrt{\frac{2g_{zz}}{g_{tt}+g_{y_1y_1}}}\right)^{-1}\\
&=\frac{\sqrt{2}}{\pi \alpha'}\left[\int^{z_{h}}_{0}\mathrm{d}z \left(b(z)z^{\frac{2}{A}}\sqrt{\frac{2}{g(z)(-f(z)+z^{2-\frac{2}{A})}}}\right)\right]^{-1}
\end{split}
\end{equation}

Fig.~\ref{figure:qhat} demonstrates the impact of spatial anisotropy on jet quenching parameters, normalized by the isotropic SYM result at zero baryon density given in Eq.~(\ref{qhatsym}). Plots (a), (b), (c), and (d) in Fig.~\ref{figure:qhat} show jet quenching parameters at different anisotropy $A=1$, $A=1.01$, $A=1.02$ and $A=1.03$ respectively. Figure (a) with $A=1$ corresponds to the isotropic case. One see that all three jet quenching parameters $\hat{q}_{(\parallel,\perp)}$, $\hat{q}_{(\perp,\parallel)}$ and $\hat{q}_{(\perp,\perp)}$  increase with anisotropic factor $A$.
And we observe the small peak around critical temperature $T_c$, which is one of the typical features of QCD phase transition. 
One reads from different anisotropic cases in Figures (b), (c), and (d) of Fig.~\ref{figure:qhat}, that in general $\hat{q}_{(\perp,\parallel)}\ge\hat{q}_{(\perp,\perp)} \ge \hat{q}_{(\parallel,\perp)}$, which indicates that the energy loss is larger in the transverse plane than along the anisotropic direction.
It also shows that the momentum broadening of an energetic parton in the anisotropic medium depends more on the direction of motion rather than the direction of momentum broadening.

\section{Conclusion }\label{chapter:summary}

The study of jet quenching properties as functions of parameters such as temperature, chemical potential, and anisotropy factor is of great relevance for understanding the anisotropic QGP. In the present work, we have taken the investigation on energy loss of a jet parton near $T_c$ at zero chemical potential under the influence of anisotropy with an EMD model.

We focus on the influences of anisotropy on several important quantities related to parton energy loss near $T_c$. It is demonstrated that with increasing anisotropic factor $A$, the drag force and jet quenching parameter go up, while the diffusion constant goes down.
The comparison of drag forces in different directions shows that energy loss near $T_c$ is larger when moving perpendicular to the anisotropy direction than parallel to the anisotropy direction.
The jet quenching parameter and diffusion constant also give the same conclusion that energy loss is stronger when the jet parton moves perpendicular to the anisotropy direction.

We also observe a peak near critical temperature $T_c$ both on the drag force ($f^{v\parallel A}$ and $f^{v\perp A}$) and the jet quenching parameter ($\hat{q}_{(\perp,\parallel)}, \, \hat{q}_{(\perp,\perp)}$, and  $\hat{q}_{(\parallel,\perp)}$) when energetic partons move nearly with the speed of light. 
However, when parton moves at a lower speed the peak of the drag force ($f^{v\parallel A}$ and $f^{v\perp A}$) near $T_c$ may disappear. Moreover, comparing numerical results of the drag force at different speeds, we see charm quark is more sensitive to the properties of the plasma than the bottom quark when the initial jet energy is fixed.

\section*{ACKNOWLEDGMENTS}
We thank Zhou-Run Zhu for his enlightening advice and very useful discussions. We also thank Zi-Qiang Zhang for his suggestions. This research is supported by the Guangdong Major Project of Basic and Applied Basic Research No. 2020B0301030008, and Natural Science Foundation of China with Project Nos. 11935007.

\bibliographystyle{unsrt}
\bibliography{ref}
\end{document}